# Production and chemical separation of $^{229}$Pa toward observation of γ rays of $^{229m}$Th

Yudai Shigekawa*, Xiaojie Yin, Akihiro Nambu, Yang Wang, Hiromitsu Haba

*Nishina Center for Accelerator-Based Science, RIKEN, Wako, Saitama 351-0198, Japan*

**Abstract**

To observe the γ rays emitted from the low-lying isomeric state of $^{229}$Th ($^{229m}$Th), we aim to dope fluoride crystals with its precursor $^{229}$Pa. In this study, we produced $^{229}$Pa by a 30-MeV proton bombardment on $^{232}$Th and developed a chemical separation method. The chemical yield of Pa was 93(4)%, and the physical production yield of $^{229}$Pa for the proton energy range of 29.0–30.1 MeV was measured to be 9.4(8) MBq/µAh, which was more than 10 times higher than those of $^{232,230,228,233}$Pa. These high chemical and production yields will allow us to prepare fluoride crystals having a sufficient amount of $^{229}$Pa for the observation of the γ rays of $^{229m}$Th.

**Introduction**

The first excited state in the $^{229}$Th nucleus, $^{229m}$Th, has an extremely low excitation energy of ~8 eV [1–4]. This low-lying isomeric state gathers much attention since laser excitation from the ground state to the isomeric state may be possible, potentially leading to an ultraprecise nuclear clock [5]. The $^{229}$Th nuclear clock is expected to have higher accuracy than current atomic clocks [6]. Moreover, the nuclear clock would be useful for testing fundamental physics [7] such as the measurement of the temporal variation of the fine structure constant [8]. The existence of $^{229m}$Th has been recently confirmed by the measurement of internal conversion (IC) electrons [9, 10]. The excitation energy of $^{229m}$Th has been measured accurately enough to start laser excitation experiments of the $^{229}$Th nucleus [1–4]. Moreover, the laser spectroscopy of the electronic states in $^{229m}$Th$^{2+}$ revealed nuclear properties of the isomer such as magnetic dipole and electric quadrupole moments [11], which are important parameters to realize the nuclear clock.

*yudai.shigekawa@riken.jp

The ultraprecise nuclear clock requires a narrow natural linewidth between the ground and isomeric states. The natural linewidth is inversely proportional to the half-life of $^{229m}$Th, based on the uncertainty principle. Hence, the IC process of $^{229m}$Th with a short half-life (7 μs [10]) must be prohibited so that $^{229m}$Th decays by the γ-ray emission with a longer half-life, which is estimated to be $10^3$–$10^4$ s [12–16]. Observing the γ rays of $^{229m}$Th and determining the radiative half-life by prohibiting the IC process are important steps toward realizing the nuclear clock. For prohibiting the IC process of $^{229m}$Th, electron binding energy should be higher than the isomer energy (~8 eV). Promising candidates are $^{229m}$Th-doped fluoride crystals, whose band gaps are calculated to be higher than 8 eV [17, 18]. Some groups prepared $^{229}$Th-doped fluoride crystals such as $^{229}$Th:LiSrAlF$_6$ [19] and $^{229}$Th:CaF$_2$ [20, 21], with which they are aiming to observe the γ rays of $^{229m}$Th by the excitation to $^{229m}$Th using synchrotron radiation [19, 22, 23]. Another method that has been tested to produce $^{229m}$Th-doped fluoride crystals is implanting $^{229m}$Th ions recoiling out of a $^{233}$U source ($T_{1/2} = 1.59 \times 10^5$ y) into fluoride crystals; however, no signals of the γ rays of $^{229m}$Th were detected with this method [24]. This is probably because $^{229m}$Th ions implanted into fluoride crystals with high-recoil energy (84 keV) were placed in unfavorable positions not surrounded by fluoride ions and thus the electron binding energy of $^{229m}$Th became lower, leading to the decay by the IC process. To overcome this problem, M. Verlinde *et al.* proposed to dope fluoride crystals with $^{229}$Ac (half-life $T_{1/2}$ = 62.7 min) [25]. First, $^{229}$Ac ions produced at the ISOLDE facility at CERN are implanted into fluoride crystals with a 30-keV energy. The crystals are then annealed to place $^{229}$Ac ions in the favorable sites surrounded by fluoride ions. Since $^{229}$Ac decays to $^{229m}$Th (branching ratio >14% [25]) by β$^-$ decay with a small recoil energy, produced $^{229m}$Th ions are also surrounded by fluoride ions, allowing to prohibit the IC process. Recently, they reported the observation of the γ rays of $^{229m}$Th by placing $^{229}$Ac into MgF$_2$ and CaF$_2$ crystals [26]. They measured the excitation energy of $^{229m}$Th to be 8.338(24) eV (148.71(42) nm) and the half-life in MgF$_2$ crystals to be 670(102) s.

We are aiming to dope fluoride crystals with $^{229}$Pa ($T_{1/2}$ = 1.50 d) instead of $^{229}$Ac for observing the γ rays of $^{229m}$Th and determining the radiative half-life, which is an important parameter to realize the nuclear clock. We implant $^{229}$Pa ions into fluoride crystals and then anneal them to place $^{229}$Pa ions in favorable sites surrounded by fluoride ions. Because $^{229}$Pa decays to $^{229m}$Th (branching ratio ~1% [27]) by the electron capture (EC) with a smaller recoil energy than the chemical binding energy, produced $^{229m}$Th ions are also surrounded by fluoride ions, and thus the γ rays of $^{229m}$Th are expected to be emitted. Although the branching ratio to $^{229m}$Th from $^{229}$Pa is smaller than that from $^{229}$Ac, the method using $^{229}$Pa has an advantage that the number of background photons originating from the Cherenkov radiation following the decay of $^{229}$Pa is extremely small. This is because the energies of most IC electrons and Auger electrons produced after the decay of $^{229}$Pa are less than 100 keV. For example, the number of electrons having higher energies than 147 keV, which is the threshold of the Cherenkov radiation for a refractive index of 1.586 ($CaF_2$, 148.7 nm) [28], is only $3\times10^{-6}$ electrons per decay of $^{229}$Pa. For $^{229}$Ac, the average energy of β$^-$ electrons is higher than 300 keV, and thus the number of the Cherenkov photons is much larger than that for $^{229}$Pa.

We estimate that ~100 kBq of $^{229}$Pa is needed to observe the γ rays of $^{229m}$Th with a simple setup consisting of photomultipliers and band-pass filters [29]. $^{229}$Pa will be ionized with the surface ionization technique [30, 31] and implanted into fluoride crystals. The ionization efficiency for Pa is reported to be 0.3–0.5% [31]; hence, 33 MBq of $^{229}$Pa is required to implant 100 kBq of $^{229}$Pa into a fluoride crystal. One of the routes to produce such a large amount of $^{229}$Pa is the proton-induced reaction on a $^{232}$Th target. The maximum cross section for the $^{232}$Th($p,4n$)$^{229}$Pa reaction was reported to be 162(14) mb at a proton energy of 29.8(2) MeV [32]. In this study, we produced $^{229}$Pa by the 30-MeV proton bombardment on the $^{232}$Th target and developed a chemical separation method for Pa to remove impurities such as the $^{232}$Th target material and fission products. The chemical purification is highly important because radioactive impurities may hinder the observation of the γ rays of $^{229m}$Th and chemical impurities may reduce the surface-ionization efficiency.

**Experimental**

*Target irradiation and γ-ray measurement*

We performed the 30-MeV proton irradiation of a $^{232}$Th target at the AVF cyclotron at RIKEN. The $^{232}$Th target consisted of two $^{232}$Th metallic foils (purity 99.5%, Goodfellow). The size and the thickness of each foil were 10×10 mm and 69.1 mg/cm$^2$, respectively. The two $^{232}$Th foils were covered with an Al foil (thickness 10 μm, purity 99.999%, Goodfellow), which were fixed to a Ta beam stopper. The $^{232}$Th target was irradiated with the proton beam collimated to a diameter of 9 mm for 590.5 min. The average beam current measured by the target working as a Faraday cup was 1.0 μA. During the irradiation, the $^{232}$Th target was cooled with water and He gas. The primary proton beam energy was measured to be 30.2 MeV with the time-of-flight method [33]. The energy of the proton bombarded on the $^{232}$Th target was calculated to be 30.1 MeV using stopping powers obtained from the SRIM code [34], considering the energy loss by a Be vacuum window (thickness 3.23 mg/cm$^2$), He gas (0.09 mg/cm$^2$), and the Al foil.

Each $^{232}$Th foil was subjected to γ-ray spectroscopy with a high-purity Ge detector 14 h after the proton irradiation. The detection efficiencies of the Ge detector were measured with an $^{152}$Eu source (Japan Radioisotope Association) and a multiple γ-ray emitting source containing $^{57,60}$Co, $^{85}$Sr, $^{88}$Y, $^{109}$Cd, $^{113}$Sn, $^{137}$Cs, $^{139}$Ce, $^{203}$Hg, and $^{241}$Am (Eckert & Ziegler Isotope Products). In this study, the radioactivities of $^{232}$Pa, $^{230}$Pa, $^{231}$Th, $^{91}$Sr, $^{93}$Y, $^{143}$Ce, $^{97}$Zr, $^{99}$Mo, $^{105}$Ru, $^{128}$Sb, $^{131m}$Te, and $^{132}$Te were determined from the γ-ray spectroscopy based on the nuclear data listed in Tables 1 and 2 in order to monitor the chemical yields of Pa, Th, and radioactive impurities during and after the chemcal separation process. The uncertainties of the radioactivities were propagated from the errors of the counting statistices, detection efficiencies, γ-ray intensities, and nuclear half-lives. Note that in the calculation of the radioactivity of $^{97}$Zr emitting the γ rays of 743.5 keV, we subtracted the interferences from the γ rays of $^{128}$Sb (743.3 keV, $I_\gamma$ = 100(7)%) and $^{131m}$Te (744.2 keV, $I_\gamma$ = 1.53(5)%) based on the other γ rays emitted from $^{128}$Sb and $^{131m}$Te. We also note that the γ-ray peaks for $^{229}$Pa and $^{228}$Pa were not observed before chemical separation due to their small γ-ray intensities (Table 1).

*Chemical separation*

The chemical separation was perfomed following the scheme shown in Fig. 1, which consists of two-step anion-exchange chromatgraphy. The first anion-exchange chromatography (Column A) was performed to separate Pa from the bulk amount of $^{232}$Th and most fission products, which is similar to the schemes reported in [35, 36]. The second anion-exchange chromatography (Column B) was developed in this study for highly selective separation of Pa, by referring to the distribution coefficients ($K_d$) of Zr, Hf, Nb, Ta, and Pa on the anion exchanger in the HCl/HF media [37]. The separation conditions with Column B were selected so as to separate Pa from Zr and Nb, which behave similarly to Pa.

The detailed procedure of the chemical separation is as follows. First, the two irradiated $^{232}$Th foils (139 mg) were dissolved with 4 mL of 11.3 M HCl and 2.1 mL of 1 M HF. After the solution was heated to dryness, 2 mL of 11.3 M HCl and 1.1 g of Al(NO$_3$)$_3$·9H$_2$O were added to the residue, where Al$^{3+}$ ions worked for masking remaining F$^-$ ions [35]. After heating the solution to dryness, the sample was dissolved with 2 mL of 11.3 M HCl, which was then poured into an strong-base anion-exchange column (Column A, Muromac® 1×8, 100-200 mesh, inner diameter (ID) 5.5 mm×height (H) 42 mm) preconditioned with 11.3 M HCl. We washed the beaker from which the sample had been moved to the column with 1 mL of 11.3 M HCl, and the washing solution was then poured into the column; this washing process was repeated twice. Next, we poured 8 mL of 11.3 M HCl into the column to elute $^{232}$Th and some fission products and then 10 mL of 6 M HCl to elute mainly Zr isotopes [35, 36]. After pouring 4 mL of 8 M HNO$_3$, Pa isotopes were eluted with 6 mL of 9 M HCl/0.1 M HF. Then, we evaporated the 9 M HCl/0.1 M HF eluate and dissolved the residue with 1 mL of 0.1 M HCl/0.1 M HF, which was then poured into an anion-exchange column (Column B, Muromac® 1×8, 100-200 mesh, ID 5.3 mm×H 23 mm) preconditioned with 0.1 M HCl/0.1 M HF. The beaker from which the sample had been moved to the column was washed with 1 mL of 0.1 M HCl/0.1 M HF, and the washing solution was poured into the column (repeated twice). After pouring 3 mL of 0.1 M HCl/0.1 M HF into the column, we poured 1 mL of 0.4 M HCl/0.1 M HF into the column six times. The fractions of 1 mL of the 0.4 M HCl/0.1 M HF eluate were separately collected (Fractions 1–6) to obtain the elution profile of Pa and other elements. We thereafter added 4 mL of 9 M HCl/0.1 M HF to the column. The γ-ray measurement was performed for every eluate with the Ge detector to determine the chemical yields for the isotopes listed in Tables 1 and 2.

The fractions 2 and 3 of the 0.4 M HCl/0.1 M HF eluate, in which purified Pa was dominantly contained, were mixed and heated to dryness. This purified Pa sample was subjected to γ-ray spectroscopy with the Ge detector in order to determine the radioactivities of the Pa isotopes and the chemical yields through the whole chemical separation process.

## Results and discussion

*Chemical separation*

The chemical yields for each eluate in the first anion exchange (Column A) are shown in Table 3. Each chemical yield was calculated by dividing the radioactivity of each isotope in each eluate by that in the $^{232}$Th target measured before the chemical separation. The chemical yield of $^{231}$Th for the 11.3 M HCl eluate was close to 100%, meaning that the bulk amount of $^{232}$Th was removed from the column. $^{91}$Sr, $^{93}$Y, and $^{143}$Ce were also efficiently removed by the elution of 11.3 M HCl. The elution of 11.3 M HCl, 6 M HCl, and 8 M HNO$_3$ removed 98(6)% of $^{97}$Zr from the column, though the small amount of $^{97}$Zr retained on the column was eluted with 9 M HCl/0.1 M HF. The elution of 4 mL of 8 M HNO$_3$ worked for removing $^{99}$Mo, $^{128}$Sb, $^{131m}$Te, and $^{132}$Te from the column. However, the removal was not complete, and these isotopes were slightly eluted with 9 M HCl/0.1 M HF. $^{232}$Pa and $^{230}$Pa were eluted with 9 M HCl/0.1 M HF with an average chemical yield of 96(6)%. Small loss of the Pa isotopes (4.2(4)% on average) was observed for the 8 M HNO$_3$ eluate, which is due to the relatively small $K_d$ value for this condition (~40 mLg$^{-1}$ [38]).

The elution profile in the second anion exchange (Column B) is shown in Fig. 2. Each chemical yield in Fig. 2 was obtained by dividing the radioactivity of each isotope in each fraction by that in the 9 M HCl/0.1 M HF eluate of Column A. First, by the loading of the sample and the elution of the 0.1 M HCl/0.1 M HF solution, 2.0(2)% of $^{232}$Pa, 63(19)% of $^{128}$Sb, and 89(8)% of $^{132}$Te were eluted from the column. Next, for the elution of 0.4 M HCl/0.1 M HF, 92(6)% of $^{232}$Pa were eluted in the first three fractions (Fractions 1–3). The γ-ray spectra for Fractions 1–3 showed the peaks only for $^{232,230,229,228,233}$Pa, and the quantities of other radioactive isotopes were lower than detection limits. We note that the detection limits of the radioactive impurities for Fractions 1–3 were relatively high due to the high radioactivity of Pa isotopes; the upper limits of the chemical yields for $^{97}$Zr, $^{99}$Mo, $^{128}$Sb, and $^{132}$Te in Fraction 1 were 0.6, 1.2, 1.0, and 0.06%, respectively, those in Fraction 2 were 12, 34, 19, and 1.3%, respectively, and those in Fraction 3 were 1.6, 4.0, 2.6, and 0.18%, respectively. However, the actual chemical yields for these radioactive impurities in Fractions 1–3 are estimated to be much lower than the above upper limits by refferring to the elution profile for Fractions 4–6 as follows. For Fraction 4, 0.02(3)% of $^{99}$Mo, 0.26(7)% of $^{128}$Sb, and 0.0030(9)% of $^{132}$Te were eluted. Because the amounts of $^{128}$Sb and $^{132}$Te in Fractions 5 and 6 are almost the same as those in Fraction 4, it is deduced that these isotopes were constantly eluted in Fractions 1–3. On the other hand, $^{99}$Mo was increasingly eluted over Fractions 4–6, meaning that the yields of $^{99}$Mo for Fractions 1–3 are estimated to be less than that for Fraction 4 (0.02(3)%). $^{97}$Zr started to be eluted from Fraction 6; hence, we can estimate that the yields of $^{97}$Zr for fractions 1–5 were less than that for Fraction 6 (0.016(16)%). These behaviours of Zr and Mo can be explained from the $K_d$ values previously reported [37, 39]. The $K_d$ value for Zr is reported to be ~100 mLg$^{-1}$ for 0.4 M HCl/0.1 M HF with the Bio-Rad AG-MP1® resin (200-400 mesh) [37], which is consistent with the little elution of Zr for 0.4 M HCl/0.1 M HF in this study. The $K_d$ value for Mo is reported to be 53 mLg$^{-1}$ for 0.2 M HCl/0.1 M HF and 30 mLg$^{-1}$ for 0.5 M HCl/0.1 M HF with the Dowex® 1×4 resin (100-200 mesh) [39]; hence, the $K_d$ value for 0.4 M HCl/0.1 M HF is roughly esitimated to be ~34 mLg$^{-1}$, which explains earlier elution of Mo than Zr in Fig. 2. Finally, 9 M HCl/0.1 M HF eluted 87(21)% of $^{97}$Zr, 5(3)% of $^{99}$Mo (5(3)%), and 3.1(11)% of $^{128}$Sb. This behavior of Zr and Mo is also consistent with the results repoted in [37, 39].

Another radioactive impurity that has not been discussed above is $^{112}$Pd ($T_{1/2}$ = 21.04 h). Although the γ rays of $^{112}$Pd (18.5 keV) could not be measured with our setup, the presence was confirmed from the γ rays of $^{112}$Ag ($T_{1/2}$ = 3.130 h), which is the daughter nuclide of $^{112}$Pd. We observed the γ rays of $^{112}$Ag in the 9 M HCl/0.1 M HF eluate of Column A. The accurate evaluation of the radioactivity of $^{112}$Pd was difficult since we did not measure the growth of $^{112}$Ag. The upper limit of the chemical yield of $^{112}$Pd in the 9 M HCl/0.1 M HF eluate of Column A was evaluated to be 5%. In the separation with Column B, the γ rays of $^{112}$Ag were observed only for the 9 M HCl/0.1 M HF eluate; hence, $^{112}$Pd is considered to be well separated from Pa isotopes. It is reported in ref. [39] that Pd is strongly adsorbed on the Dowex® 1×4 resin (100-200 mesh) for 0.2 M HCl/0.1 M HF and 0.5 M HCl/0.1 M HF while the adsorbption is much weaker for 9 M HCl/0.1 M HF ($K_d$ = 35 mLg$^{-1}$), which is consistent with our experimental result. We estimate based on [39] that the elution of Pd with 0.4 M HCl/0.1 M HF is later than Zr. Hence, the chemical yields of $^{112}$Pd in Fractions 1–6 of the 0.4 M HCl/0.1 M HF eluate are estimated to be less than 0.016(16)%.

We took Fractions 2 and 3 containing dominant amounts of Pa as the purified Pa smaple. The γ-ray spectrum of the purified Pa sample is shown in Fig. 3, from which the radioactivities of $^{232}$Pa and $^{230}$Pa were determined. The whole chemical yields of $^{232}$Pa and $^{230}$Pa were calculated by dividing each radioactivity in the purified Pa sample by that in the $^{232}$Th target; the yields were 92(5) % for $^{232}$Pa and 96(7)% for $^{230}$Pa. By taking a weighed average, the whole chemical yield of Pa was determined to be 93(4)%. The whole chemical yields of radioactive impurities were esmimated to be <3×10$^{-4}$% for $^{97}$Zr, <1×10$^{-3}$% for $^{99}$Mo, 1.1(2)×10$^{-2}$% for $^{128}$Sb, 2.4(2)×10$^{-4}$% for $^{131m,132}$Te, and <3×10$^{-3}$% for $^{112}$Pd. Their radioactivities at the end of bombardment (EOB) were estimated to be <100 Bq for $^{97}$Zr, <120 Bq for $^{99}$Mo, 1600(600) Bq for $^{128}$Sb, 16(5) Bq for $^{131m}$Te, 12(4) Bq for $^{132}$Te, and <900 Bq for $^{112}$Pd (for simplicity, short-lived mother nuclides of these isotopes were assumed to completely decay to these isotopes at the EOB). These radioactivities were much lower than those of $^{232}$Pa (5.34(17) MBq at the EOB) and $^{230}$Pa (2.26(10) MBq at the EOB). In the future experiments to measure the 8-eV γ rays from $^{229m}$Th, Pa isotopic impurities will be the main source of background photons that may hinder the observation of the $^{229m}$Th γ rays, and the radioactive impurities of other elements having much lower radioactivities than Pa isotopes will be negligible as the source of the background photons.

*Production yields and cross sections of Pa isotopes*

First, the physical production yields and cross sections of $^{232}$Pa and $^{230}$Pa were calculated from their radioactivities in the $^{232}$Th target measured before the chemical separation. The results are shown in Table 4, where we used a beam intensity of 1.0(3) μA and a target thickness of 138(3) mg/cm$^2$ for the calculation. The obtained cross section of $^{232}$Pa (13.1(8) mb) at the proton energy range of 29.0–30.1 MeV is consistent with the recently measured values of 13.0(10) and 13.1(10) mb at the proton energies of 29.6±0.5 and 29.8±0.2 MeV, respectively [32]. The cross section of $^{230}$Pa (64(5) mb) measured in this study is also consistent with 59.4(43) and 59.8(45) mb at the proton energies of 29.6±0.5 and 29.8±0.2 MeV, respectively [32]. Very recent paper reports the cross sections of 66.9(47) and 61.2(43) mb at the proton energies of 28.50±0.45 and 30.74±0.42 MeV, respectively [40], which are also consistent with our measurement.

Next, from the γ-ray spectrum of the purified Pa sample (Fig. 3), we calculated the radioactivities of $^{229}$Pa and $^{233}$Pa to be 80(5) and 0.072(4) MBq at the EOB, respectively. Since two different half-lives have been reported for $^{228}$Pa ($T_{1/2}$ = 22(1) h [41] and $T_{1/2}$ = 19.5(4) h [42]), the radioactivity of $^{228}$Pa at the EOB was calculated for each half-life: 5.4(4) and 6.5(4) MBq for the half-lives of 22(1) and 19.5(4) h, respectively. The whole chemical yield of 93(4)% determined from $^{232}$Pa and $^{230}$Pa allowed us to calculate the radioactivities of $^{229}$Pa, $^{228}$Pa, and $^{233}$Pa contained in the $^{232}$Th target before the chemical separation, from which we obtained the physical production yields and cross sections of these isotopes as listed in Table 4. The cross section of $^{229}$Pa at the proton energy range of 29.0–30.1 MeV was determined to be 219(20) mb, 35% higher than that the reported value of 162(14) mb [32] at the proton energy of 29.8±0.2 MeV, which was the peak position of the excitation function. Further measurements are required to determine the cross sections for the $^{232}$Th($p,4n$)$^{229}$Pa reaction more accurately. Next, the cross section of $^{228}$Pa measured in this study was 9.5(9) mb for $T_{1/2}$ = 22(1) h and 10.4(9) mb for $T_{1/2}$ = 19.5(4) h, which are much higher than the reported values of 3.8(3) and 3.9(5) mb at the proton energies of 29.6±0.5 and 29.8±0.2 MeV, respectively [32]. According to the results of the theoretical model calculation in the TENDL-2021 library [43], the cross sections in the $^{232}$Th($p,5n$)$^{228}$Pa reaction at proton energies of 28.0, 30.0, and 35.0 MeV are calculated to be 0.00014, 0.11, and 20 mb, respectively; hence, the cross section at around 30 MeV is estimated to steeply increase with increasing the proton energy. The proton energy bombarded on the $^{232}$Th target in this study (30.1 MeV) is higher than those for the previous study (29.6±0.5 and 29.8±0.2 MeV [32]), which may lead to the higher cross section measured in this study than those in the previous study due to the steep increase of the cross section at around 30 MeV.

We determined the physical production yield of $^{229}$Pa to be 9.4(8) MBq/µAh for the proton energy range of 29.0–30.1 MeV, which is >10 times higher than the yields of other Pa isotopes (Table 4). Therefore, we confirmed that the 30-MeV proton bombardment on $^{232}$Th is an ideal condition to obtain a $^{229}$Pa-rich sample. If we irradiate a $^{232}$Th metallic target (138 mg/cm$^2$) with a 30-MeV proton beam at a current of 10 µA for 1.5 d and then perform the chemical separation with the scheme developed in this study, we will be able to obtain 1550(120) MBq of $^{229}$Pa 20 h after the end of the irradiation (reasonable time for target cooling and chemical separation). This radioactivity is much higher than 33 MBq, which is required to implant 100 kBq of $^{229}$Pa into a fluoride crystal with the surface ionization technique as mentioned in Introduction. Hence, the sufficient quantity of $^{229}$Pa to observe the γ rays of $^{229m}$Th can be produced with the method developed in this study. On the other hand, the impurities of Pa isotopes produced on the above condition will be 95(4) MBq of $^{232}$Pa, 77(4) MBq of $^{230}$Pa, 72(7) MBq of $^{228}$Pa ($T_{1/2}$ = 22(1) h), and 2.5(2) MBq of $^{233}$Pa. Among these isotopes, $^{232}$Pa and $^{230}$Pa, which emit β$^-$ electrons, will be the main source of the Cherenkov photons and may interfere with the observation of the γ rays of $^{229m}$Th. However, the situation is quite better than the case for $^{229}$Ac that emits β$^-$ electrons with an average energy of 387 keV because the average energies of β$^-$ electrons for $^{232}$Pa and $^{230}$Pa are lower (~90 keV for $^{232}$Pa and ~150 keV for $^{230}$Pa) and the branching ratio of the β$^-$ decay of $^{230}$Pa is small (7.8%) [44]. Moreover, we will be able to reduce the amounts of the Pa impurities by performing mass separation of $^{229}$Pa ions produced by surface ionization, which will allow the low-background measurement of the γ rays of $^{229m}$Th, compared with the measurement using $^{229}$Ac.

**Conclusions**

We are aiming to observe the γ rays of $^{229m}$Th and determine its radiative half-life by preparing $^{229}$Pa-doped fluoride crystals. Toward this goal, we produced $^{229}$Pa in the $^{232}$Th(*p*,4*n*)$^{229}$Pa reaction with the 30-MeV proton beam and developed the chemical separation method of Pa from the $^{232}$Th target and fission products, consisting of two-step anion-exchange chromatography. The first anion-exchange chromatography allowed the separation of Pa isotopes from $^{232}$Th and some fission products (e.g. $^{91}$Sr, $^{93}$Y, $^{143}$Ce, and $^{105}$Ru) with a chemical yield of 96(6)%. The second anion-exchange chromatography was performed for reducing the radioactive impurities of $^{97}$Zr, $^{99}$Mo, $^{128}$Sb, $^{131m}$Te, $^{132}$Te, and $^{112}$Pd. We found that the separation scheme with the 0.1 M HCl/0.1 M HF and 0.4 M HCl/0.1 M HF solutions works well for selectively separating Pa from other elements; the radioactive impurities except for Pa isotopes ($^{232,230,228,233}$Pa) were reduced to less than detection limits of the γ-ray measurement. The chemical yield of Pa through the entire chemcal separation process was 93(4)%. The cross section and production yield of $^{229}$Pa for the proton energy range of 29.0–30.1 MeV were measured to be 219(20) mb and 9.4(8) MBq/μAh, respectively. The production yields of the isotopic impurities ($^{232,230,228,233}$Pa) were found to be >10 times less than that of $^{229}$Pa. We will be able to obtain 1550(120) MBq of $^{229}$Pa by the 1.5-d irradiation of a $^{232}$Th target (138 mg/cm$^2$) with a 30-MeV proton beam at a current of 10 μA (cooling time 20 h), which is high enough to dope fluoride crystals with 100 kBq of $^{229}$Pa by surface ionization and ion implantation into the crystals. We are now developing the ionization and implantation system for $^{229}$Pa [45] and the photon measurement apparatus for observing the γ rays of $^{229m}$Th [29]. We will soon perform the experiments to make $^{229}$Pa-doped fluoride crystals and measure the γ rays of $^{229m}$Th.

**Acknowledgements**

This experiment was performed at RI Beam Factory operated by RIKEN Nishina Center and CNS, University of Tokyo. This work was supported by JSPS KAKENHI Grant Number 19K23445.

Tables

**Table 1** Half-life, γ-ray energy ($E_\gamma$), and γ-ray intensity per decay ($I_\gamma$) for Pa isotopes observed in this study. Nuclear reactions to produce these isotopes are also shown. The isotope data is retrieved from [44] unless otherwise noted.

| Nuclide | Half-life | $E_\gamma$ (keV) | $I_\gamma$ (%) | Reaction |
|---|---|---|---|---|
| $^{232}$Pa | 1.31(2) d | 969.3 | 42.3(6) | $^{232}$Th($p,n$) |
| | | 894.4 | 19.6(3) | |
| | | 387.9 | 6.55(15) | |
| $^{230}$Pa | 17.4(5) d | 951.9 | 29.6(18) | $^{232}$Th($p,3n$) |
| | | 918.5 | 8.3(4) | |
| | | 898.7 | 5.8(4) | |
| $^{229}$Pa | 1.50(5) d | 119.0 | 0.129(7) | $^{232}$Th($p,4n$) |
| | 1.55(1) d[a] | 117.2 | 0.047(3) | |
| | 1.5(1) d[b] | | | |
| $^{228}$Pa | 22(1) h[c] | 911.2[c] | 2.26(16)[c] | $^{232}$Th($p,5n$) |
| | 19.5(4) h[b] | 964.8[c] | 1.12(9)[c] | |
| | | 409.4[c] | 0.93(7)[c] | |
| | | 129.1[c] | 0.43(3)[c] | |
| $^{233}$Pa | 26.975(13) d | 311.9 | 38.2(4) | $^{232}$Th($p,\gamma$) |
| | | 300.1 | 6.54(8) | $^{232}$Th($n,\gamma$)$^{233}$Th |

[a]From ref. [32]
[b]From ref. [42]
[c]From ref. [41]

**Table 2** Half-life, γ-ray energy ($E_\gamma$), and γ-ray intensity per decay ($I_\gamma$) for $^{231}$Th and fission products quantified in this study. The nuclide data is retrieved from [44].

| Nuclide | Half-life | $E_\gamma$ (keV) | $I_\gamma$ (%) |
|---|---|---|---|
| $^{231}$Th | 25.57(8) h | 84.2 | 6.8 |
| $^{91}$Sr | 9.65(6) h | 1024.3 | 33.5 |
| $^{93}$Y | 10.18(8) h | 266.9 | 7.4(11) |
| $^{143}$Ce | 33.039(6) h | 293.3 | 42.8(4) |
| $^{97}$Zr | 16.749(8) h | 743.4 | 93.09 |
| $^{99}$Mo | 65.924(6) h | 739.5 | 12.2 |
| | | 181.1 | 6.05(12) |
| $^{105}$Ru | 4.439(11) h | 676.4 | 15.82(2) |
| $^{128}$Sb | 9.05(4) h | 754.0 | 100(7) |
| | | 314.1 | 61(4) |
| | | 636.2 | 36(3) |
| $^{131m}$Te | 33.25(25) h | 852.2 | 19.9(5) |
| | | 793.7 | 13.4(3) |
| | | 1206.6 | 9.41(22) |
| $^{132}$Te | 3.204(13) d | 228.2 | 88(3) |

**Table 3** Chemical yields for the first anion-exchange chromatography (Column A). The cells shown as "−" mean that the corresponding isotopes were under the detection limits in the γ-ray measurements. The chemical yields were calculated by dividing the radioactivities in each eluate by those in the $^{232}$Th target.

| Nuclide | Chemical yield (%) | | | |
|---|---|---|---|---|
| | 11.3 M HCl | 6 M HCl | 8 M HNO$_3$ | 9 M HCl/0.1 M HCl |
| $^{232}$Pa | − | − | 4.9(7) | 96(7) |
| $^{230}$Pa | − | − | 3.7(6) | 96(10) |
| $^{231}$Th | 103(6) | 0.33(13) | − | − |
| $^{91}$Sr | 100(6) | 0.50(9) | − | − |
| $^{93}$Y | 114(22) | − | − | − |
| $^{143}$Ce | 102(6) | 0.3(2) | − | − |
| $^{97}$Zr | 79(5) | 17.6(12) | 0.8(2) | 0.51(12) |
| $^{99}$Mo | − | − | 103(7) | 1.2(6) |
| $^{105}$Ru | 35(9) | 14(2) | 57(6) | − |
| $^{128}$Sb | − | − | 7.6(9) | 2.1(5) |
| $^{131m}$Te | − | − | 91(7) | 3.4(8) |
| $^{132}$Te | − | − | 91(7) | 4.1(3) |

**Table 4** Physical production yields and cross sections of Pa isotopes for the proton energy range of 29.0–30.1 MeV.

| Nuclide | Physical production yield (MBq/µAh) | Cross section (mb) |
| --- | --- | --- |
| $^{229}$Pa | 9.4(8)[a] | 219(20)[a] |
| $^{232}$Pa | 0.65(3) | 13.1(8) |
| $^{230}$Pa | 0.238(17) | 64(5) |
| $^{228}$Pa | 0.67(6)[b] | 9.5(9)[b] |
|  | 0.83(7)[c] | 10.4(9)[c] |
| $^{233}$Pa | 0.0078(6) | 3.2(3) |

[a]Calculated with $T_{1/2}$ = 1.50(5) d [44]

[b]Calculated with $T_{1/2}$ = 22(1) h [41]

[c]Calculated with $T_{1/2}$ = 19.5(4) h [42]

Figures

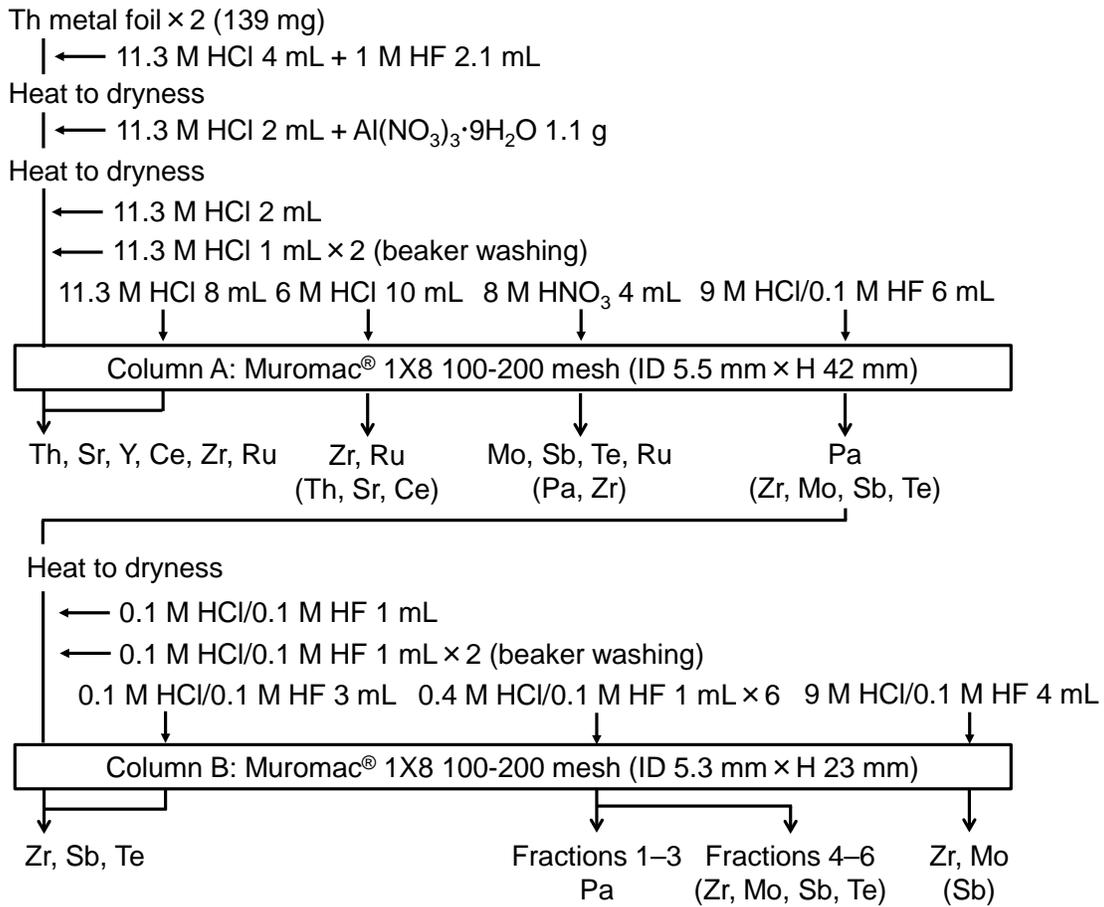

**Fig. 1** Scheme of the chemical separation performed in this study. Elements observed with γ-ray spectroscopy are shown for each eluate (the elements whose chemical yields were below 5% are placed in parentheses).

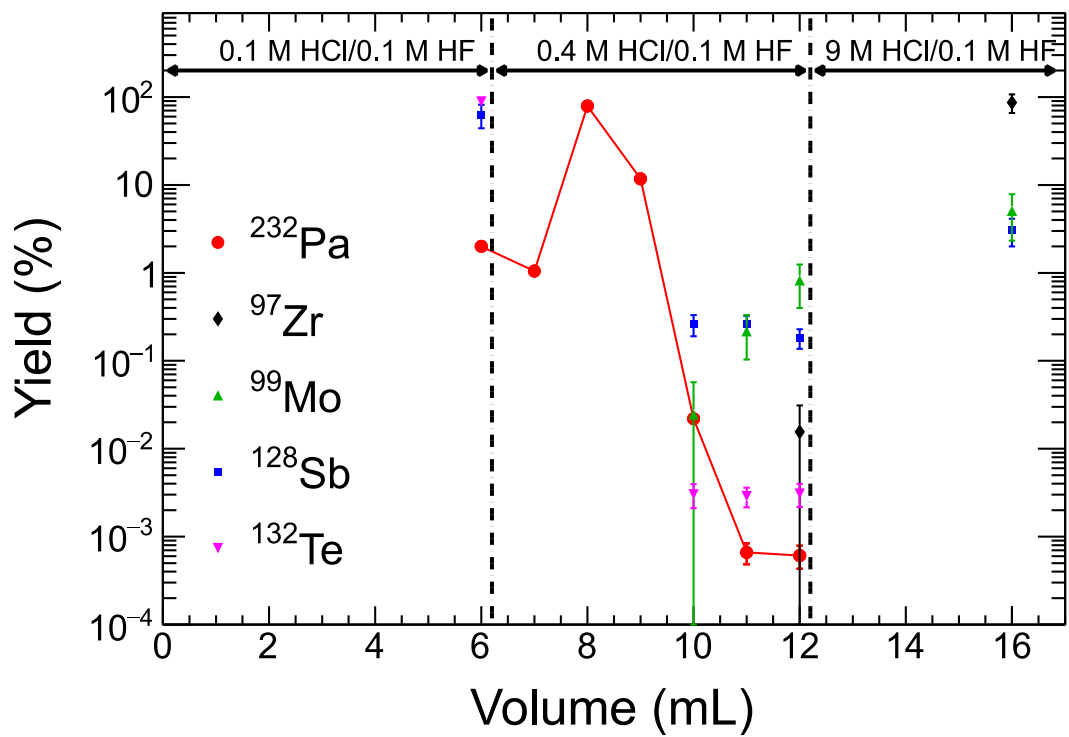

**Fig. 2** Elution profile for the second anion-exchange chromatography (Column B). The chemical yield of each nuclide in each fraction was calculated by dividing its radioactivity in each fraction by that in the 9 M HCl/0.1 M HCl eluate of Column A. The yields for $^{230}$Pa and $^{131m}$Te were consisitent with those for $^{232}$Pa and $^{132}$Te, respectively, and thus they are not plotted for visibility. The line for $^{232}$Pa is only a guide to the eye.

**Fig. 3** γ-ray spectrum of the purified Pa sample prepared by mixing Fractions 2 and 3(a: $^{232}$Pa, b: $^{230}$Pa, c: $^{229}$Pa, d: $^{228}$Pa, e: $^{233}$Pa, x: x-rays, g: background).